\documentstyle{article}
\title{\vspace*{-1.cm}\hspace*{9.6cm} {\large gr-qc/9504006}
\vspace*{.7cm}\\
\Large{{\bf Canonical Quantization of Cylindrically Symmetric Models
\vspace*{.7cm}}}}
\author{
Guillermo A. Mena Marug\'an\vspace*{.4cm}\\
Instituto de Matem\'aticas y F\'{\i}sica
Fundamental, C.S.I.C.,\\ Serrano 121, 28006 Madrid, Spain.}
\date{\empty}

\begin{document}

\maketitle
\large
\setlength{\baselineskip}{.825cm}
\vspace*{.15cm}

\begin{center}
{\bf Abstract}
\end{center}
\vspace*{.4cm}

We carry out the canonical quantization of the Levi-Civit\`a family
of static and cylindrical solutions. The reduced phase space
of this family of metrics is proved to coincide with that
corresponding to the Kasner model, including the associated
symplectic structures, except for that the respective domains of
definition of one of the phase space variables are not identical.
Using this result, we are able to construct a quantum model that
describes spacetimes of both the Levi-Civit\`a and the Kasner type,
and in which the three-dimensional spatial topology is not uniquely fixed.
Finally, we quantize to completion the subfamily of
Levi-Civit\`a solutions which represent the exterior gravitational field
of a straight cosmic string. These solutions are conical geometries,
ie, Minkowski spacetime minus a wedge. The quantum theory obtained provides
us with predicitions about the angular size of this wedge.

\vspace*{1.cm}


\newpage
\renewcommand{\thesection}{\Roman{section}.}
\renewcommand{\theequation}{\arabic{section}.\arabic{equation}}
\renewcommand{\thefootnote}{a}

\section {Introduction}

Spacetimes with cylindrical symmetry are employed in General Relativity
to describe a variety of physically interesting situations. The general form
of the vacuum four-dimensional metric for cylindrically symmetric spacetimes
was first obtained by Kompaneets.$^1$ Particular examples of such a
metric provide exact solutions to Einstein's equations which represent
cylindrical gravitational waves.$^{2-4}$ These waves can be superimposed
on the exterior gravitational field created by a massive static cylinder.$^2$
Furthermore, the existence of cylindrical symmetry allows one to define a
locally measurable gravitational energy, which can be proved to be localized
for propagating cylindrical waves.$^3$

Cylindrical symmetry, on the other hand, guarantees that the spacetime
pos\-sesses two spacelike Killing vectors.$^5$ Thanks to this fact, once
a cylindrical solution is known, one can apply to it the generalized
soliton transformation of Belinski\v{\i} and Zakharov$^6$ to generate
new vacuum solutions. The cylindrical metrics which have been more
frequently used as seeds for the Belinski\v{\i}-Zakharov technique are the
Levi-Civit\`a (LC from now on) family of solutions.$^{7,8}$
In this way, it has been possible to construct a series
of cylindrical solutions which describe the
propagation of gravitational waves on a LC background.$^5$

The LC metrics are the most general static and cylindrically symmetric
solution to Einstein's equations in vacuum, and can be expressed in the form
\begin{equation} ds^2=C\;\rho^{2p^2-\frac{1}{2}}(-dt^2+d\rho^2)
+AB\;\rho^{1+2p}\;d\phi^2+\frac{A}{B}\,\rho^{1-2p}\;dz^2,\end{equation}
where $A$, $B$ and $C$ are three positive constants and $p$ a real constant.
The time coordinate $t$ and the axial coordinate $z$ take on real values,
$\rho\in I\!\!\!\,R^+$ is the radial coordinate, and $\phi\in S^1$ the
polar angle.

Except when $p=\frac{1}{2}$, metrics (1.1) can be interpreted as the
exterior gravitational field produced by an static cylinder.$^{2,7}$
For cylinders made of ordinary matter, ie, with positive mass, we have
$p>\frac{1}{2}\;$.$^2$ When $p<\frac{1}{2}$, the
mass of the cylinder must be negative, and the gravitational field described
by metric (1.1) can be regarded as repulsive.$^4$
All metrics (1.1) with
$p\neq \frac{1}{2}$ present a naked singularity at $\rho=0$. This singularity
can be reinterpreted as a line source, which can be represented by an
appropriate stress-energy-momentum tensor on the axis $\rho=0\;$.$^9$

When $p=\frac{1}{2}$, a suitable rescaling of the coordinates
$t$, $\rho$ and $z$, which leaves invariant their respective domains of
definition, leads to the subfamily of metrics
\begin{equation} ds^2=-dt^2+d\rho^2
+\,\frac{AB}{C}\,\rho^2\,d\phi^2+dz^2.\end{equation}
For $C=AB$ this is the metric of Minkowski spacetime in cylindrical
coordinates. Of much more physical interest, nevertheless,
is the case $C>AB$. Metric (1.2)
describes then a conical spacetime, ie, flat spacetime minus a wedge.$^{10}$
The angular size of this wedge is
\begin{equation} \Delta=\;2\pi\left(1-\sqrt{\frac{AB}{C}}\right).
\end{equation}
Such a conical geometry can actually be interpreted as the exterior field
of a uniform, static and infinite straight cosmic string with linear
energy density equal to $2\Delta$ (we use units where $16\pi G=c=
\hbar=1$).$^{10,11}$ This kind of topological defects$^{12}$ could play a
relevant role in cosmology,$^{10}$ as they could serve
as seeds for eternal inflation,$^{13}$ and provide the density
fluctuations necessary for galaxy formation.$^{14}$ Cosmics strings would
also have observational gravitational effects if they exist. They should
produce a steplike discontinuity in the microwave background,$^{15}$ and act
as gravitational lenses, forming double images of astrophys\-ical
objects.$^{16}$ In the case of a straight cosmic string, for instance, the
angular separation between these double images would be proportional to the
deficit angle $\Delta$ of the exterior conical geometry.$^{10}$

In this work, we will construct a quantum theory for the
description of the cylindrically symmetric
LC spacetimes, and apply it to the study of different physical situations.
In order to achieve this goal, we will follow the extension of Dirac's
canonical quantization programme$^{17}$ ellaborated by Ashtekar
{\it et al}.$^{18}$ This extended programme has already proved extremely
successful in dealing with the quantization of gravitational minisuperspace
models.$^{19-21}$
In the case of the LC spacetimes, the quantization procedure to be employed
will consist of the following steps. We will first find the symplectic
structure of the associated reduced phase space. In this
space, we will choose a complete set of elementary observables which form
a Lie algebra under the Poisson brackets determined by the symplectic
structure. These observables will be represented as operators acting
on a vector space, in such a way that their commutators reproduce the
corresponding classical Poisson brackets. We will then select the inner
product in the representation constructed by imposing a set of reality
conditions,$^{18,22}$ namely, by promoting the complex conjugation relations
between classical elementary observables to adjointness requirements on
quantum operators.

This paper is organized as follows.
Section II is devoted to the canonical quantization
of the LC metrics. In Section III we construct a quantum model that describes
the LC and the Kasner$^8$ families of solutions altogether.
The quantization of the subfamily of LC metrics that
represent the exterior field of a straight cosmic string is presented
in Section IV. We comment there on the type of predictions
that can be extracted from the quantum theory obtained.
Finally, we summarize and conclude in Section V.

\section{Quantum Analysis of the LC Solutions}
\setcounter{equation}{0}

The LC metrics (1.1) provide the most general static, cylindrical
solution to Ein\-stein's equations in vacuum. We can regard these
metrics as the classical solutions of a cylindrically symmetric
minisuperspace model in which all metric functions, and thus all degrees
of freedom, depend exclusively on the radial coordinate $\rho$, and not on
time. The Hamiltonian of such a minisuperspace model, obtained as a linear
combination of all the first-class constraints of the system,$^{17}$ must
then dictate the evolution of the metric functions in the radial coordinate,
rather than generating a true time evolution. In spite of this peculiarity,
we will show that the quantization of the LC spacetimes poses no special
difficulty if one interprets all gravitational variables as
complex-valued functions on phase space, as proposed by Ashtekar.$^{18}$

In order to quantize the LC family of solutions, we will first prove that,
with a reinterpretation of the spacetime coordinates, the LC and the Kasner
metrics coincide except for the sign of the parameter $C$ appearing in
Eq. (1.1). The Kasner solutions (ie, the Bianchi type I diagonal metrics)
can be written in the form$^{20}$
\begin{equation} ds^2=\tilde{C}\,\tilde{t}^{2p^2-\frac{1}{2}}
(-d\tilde{t}^2+d\tilde{\rho}^2)+AB\,\tilde{t}^{1+2p}\,d\phi^2
+\frac{A}{B}\,\tilde{t}^{1-2p}\,dz^2,\end{equation}
being $\tilde{t}\in I\!\!\!\,R^+$ the time coordinate, $(\tilde{\rho},\phi,
z)$ a set of real spatial coordinates, $A$, $B$ and $\tilde{C}$ positive
constants, and $p$, in principle, a real constant.
Let us assume that, in the above equation,
$\tilde{\rho},z\in I\!\!\!\,R$ and $\phi\in S^1$. Identifying the
coordinates $\rho$ and $t$ of Eq. (1.1) with $\tilde{t}$ and $\tilde{\rho}$,
respectively (recall that $\rho,\tilde{t}\in I\!\!\!\,R^+$ and $t,
\tilde{\rho}\in I\!\!\!\,R$), it is now obvious that the LC solutions adopt
the same expression as the Kasner metrics (2.1), the only difference being
that $\tilde{C}=-C<0$ in the LC case, whereas $\tilde{C}>0$ for the Kasner
spacetimes.

As a consequence, the quantization of the LC solutions
can be made by quantizing the Kasner metrics (2.1), provided that one
replaces the non-holonomic constraint $\tilde{C}>0$ with $\tilde{C}<0$.
In fact, the Kasner solutions have already been quantized.$^{20,21}$
Reference 20 contains a detailed analysis of the
canonical quantization of these metrics which employs the same kind of
notation and mathematical language used in this work. In particular, it
was shown that the reduced phase space of the Kasner model has a symplectic
structure given by
\begin{equation} \Gamma=\;dA\wedge dP_A\;+\;dp\wedge dP_p\;,\end{equation}
where
\begin{equation} P_A=p\ln{B}-\frac{1}{2}\ln{\tilde{C}}\;,\;\;\;\;\;\;
P_p=A\;(\ln{B}-2p)\;.\end{equation}
It was also argued$^{20}$ that, as long as the spatial coordinates
$\phi$ and $z$ are physically undistinguishable, one should restrict the
parameter $p$ to be positive so that each possible 4-geometry described
by a metric of the form (2.1) is taken into account only once. In our case,
however, the symmetries and domains of definition of $\phi$ and $z$ are
clearly different. So, we must let $p$ run over the whole real axis if
all possible geometries are to be considered.

We thus have that, for the Kasner metrics, $p\in I\!\!\!\,R$ and
$A,B,\tilde{C}\in I\!\!\!\,R^+$. Then, from Eq. (2.3),
\begin{equation} P_A,P_p\in I\!\!\!\,R\;.\end{equation}
The symplectic structure of the reduced phase space determined by Eq. (2.2)
can therefore be interpreted as that corresponding to the cotangent bundle
over $I\!\!\!\,R^+\times I\!\!\!\,R$.

For the LC metrics, on the other hand, $p$, $A$ and $B$ take on the same
ranges of values as in the Kasner model, but now $\tilde{C}\in I\!\!\!\,R^-$.
Hence, although we still get
\begin{equation} A\in I\!\!\!\,R^+\;\;\;\;{\rm and}\;\;\;\; p,P_p\in
I\!\!\!\,R\;,\end{equation}
the variable $P_A$ becomes complex:
\begin{equation} P_A=p\ln{B}-\frac{1}{2}\ln{|\tilde{C}|}-\frac{\pi}{2}i
\;\in I\!\!\!\,R-\frac{\pi}{2}i\;,\end{equation}
where we have taken $\ln{(-1)}=\pi i$. Notice that, being the imaginary
part of $P_A$ constant, one can always absorb it by shifting the origin
of $P_A$ in the complex plane. In this way, instead of working
with $P_A$ complex for the LC metrics, one can simplify calculations by
introducing the new variable
\begin{equation} \tilde{P}_A= P_A +\frac{\pi}{2}i=p\ln{B}-\frac{1}{2}
\ln{|\tilde{C}|}\;,\end{equation}
which is real in the analysed case,
\begin{equation} \tilde{P}_A\in I\!\!\!\,R\;.\end{equation}
The symplectic form (2.2) can now be rewritten
\begin{equation} \Gamma=\;dA\wedge d\tilde{P}_A\;
+\;dp\wedge dP_p\;.\end{equation}
{}From Eqs. (2.5) and (2.8), we then conclude that, for the LC spacetimes,
the symplectic structure of the reduced phase space can be identified with
that of the cotangent bundle over $I\!\!\!\,R^+\times I\!\!\!\,R$. This is
precisely the symplectic structure obtained for the Kasner model.
The only physical difference is that, in terms of the reduced phase space
variables employed, the parameter $\tilde{C}$ of Eq. (2.1)
adopts the following expressions for the Kasner and
the LC solutions:
\begin{equation} \tilde{C}=\exp{\left(\!4p^2\!+2\frac{pP_p}{A}\!-
2P_A\!\right)}
\;,\;\;\;\;\;\;\;\;\;\tilde{C}=-\exp{\left(\!4p^2\!+2\frac{pP_p}{A}\!
-2\tilde{P}_A\!\right)}\;,\end{equation}
respectively.

A complete set of elementary variables in the reduced phase space of
the LC metrics is formed by $p$, $P_p$, $A$ and the generalized
momentum
\begin{equation} L_A=\;A\;\tilde{P}_A.\end{equation}
Since these are reduced phase space variables, they all are gravitational
observables. Besides, they form a Lie algebra under Poisson brackets,
the only non-vanishing brackets being
\begin{equation} \{A,L_A\}=A\;,\;\;\;\;\;\;\;\;\{p,P_p\}=1\;.\end{equation}

To quantize the system, we will represent these elementary observables as
operators acting on the vector space of complex functions $\Psi(A,p)$ over
$I\!\!\!\,R^+\times I\!\!\!\,R$, each function $\Psi(A,p)$ corresponding to
a quantum state. The action of such operators will be given by
\begin{equation} \hat{A} \Psi(A,p)=A\Psi(A,p)\;,\;\;\;\;\;
\hat{L}_A\Psi(A,p)=-iA\partial_A\Psi(A,p)\;,\end{equation}
\begin{equation} \hat{p} \Psi(A,p)=p\Psi(A,p)\;,\;\;\;\;\;
\hat{P}_p\Psi(A,p)=-i\partial_p\Psi(A,p)\;.\end{equation}
These definitions guarantee that the introduced operators form the same
algebra under commutators as their classical counterparts under Poisson
brackets.

We can now fix the inner product in the representation constructed by
promoting the reality conditions on the elementary observables to
adjointness requirements on quantum operators.$^{22}$
Since the chosen elementary observables are real, we must demand that the
operators (2.13,14) be self-adjoint. These hermiticity conditions determine
the unique inner product (up to a positive global factor)
\begin{equation} <\Phi,\Psi>=\int_{I\!\!\!\,R^+} \frac{dA}{A}
\int_{I\!\!\!\,R} dp \;\Phi^{\ast}(A,p)\;\Psi(A,p)\;,\end{equation}
where $\Phi^{\ast}$ is the complex conjugate to $\Phi$. The Hilbert space
of physical states is thus $L^2(I\!\!\!\,R^+\times I\!\!\!\,R, dAdp/A)$.
This completes the quantization of the LC solutions.

\section{A Quantum Model with Non-Fixed 3-Topology}
\setcounter{equation}{0}

In this section, we will elaborate a quantum model which could describe
both the LC and the Kasner families of solutions as a whole. Being the
topology of the surfaces of constant time for these two types of
classical spacetimes different, we notice that the model to be quantized
should not possess a uniquely fixed spatial topology.

We recall that both the Kasner and the LC solutions can be expressed in the
form (2.1), with $A,B\in I\!\!\!\,R^+$ and $p\in I\!\!\!\,R$. Metrics (2.1)
correspond to Kasner or LC solutions, respectively, depending on whether
$\tilde{C}$ is positive or negative. In both cases, the symplectic form
on the reduced phase space is provided by formulae (2.2,3). Suppose then
that one performs a time reversal on the Kasner spacetimes. This time
reversal can be seen to change the global sign of the symplectic form (2.2),
although it does not alter the four-dimensional geometry. In fact, the same
flip of sign can be obtained in Eq. (2.2) by means of the transformation
$A\rightarrow -A$ and $B\rightarrow -B$, which leaves invariant metrics
(2.1). Hence, instead of using Eq. (2.2), we can equivalently employ
the following symplectic form for the Kasner model
\begin{equation} \Gamma=-\;(dA\wedge dP_A\;+\;dp\wedge dP_p)\;.\end{equation}

Let us define now
\begin{equation} \tilde{A}=\epsilon\;A\;,\;\;\;\;\;\;\;\;\;\;
\tilde{B}=\epsilon\;B\;,\end{equation}
\begin{equation} R=-\frac{\tilde{C}}{\tilde{A}}\;,\;\;\;\;\;\;\;\;\;\;
S=\,\frac{\tilde{A}}{\tilde{B}}\;,\end{equation}
where $\epsilon=1$ for the LC spacetimes, and $\epsilon=-1$ for the Kasner
solutions. Then, metrics (2.1) adopt the expression
\begin{equation} ds^2=-\tilde{A}R\;\tilde{t}^{2p^2-\frac{1}{2}}
(-d\tilde{t}^2+d\tilde{\rho}^2)+\frac{\tilde{A}^2}{S}\;\tilde{t}^{1+2p}
\;d\phi^2+S\;\tilde{t}^{1-2p}\;dz^2\;,\end{equation}
with
\begin{equation} p\in I\!\!\!\,R\;,\;\;\;\;\;\;
R,S\in I\!\!\!\,R^+\end{equation}
and
\begin{equation} \tilde{A}\in\left\{\begin{array}{lc} I\!\!\!\,R^+
& {\rm for\;\; LC\;\; solutions}\\ I\!\!\!\,R^- &
{\rm for\;\; Kasner\;\; solutions.}\end{array}\right.\end{equation}
On the other hand, making use of relations (2.3) and (2.7), it is not
difficult to check that the symplectic forms (3.1) and (2.9), valid
on the respective reduced phase spaces of the Kasner and the LC models,
can both be rewritten as
\begin{equation} \Gamma=\;d\tilde{A}\wedge d\Pi_{\tilde{A}}\;+\;
dp\wedge d\Pi_p\;,\end{equation}
where the momenta $\Pi_{\tilde{A}}$ and $\Pi_p$ are
\begin{equation} \Pi_{\tilde{A}}=-\,p\ln{S}-\frac{1}{2}\ln{R}\;,\;\;\,
\;\;\;\Pi_p=-\,\tilde{A}\;(\ln{S}+2p-1)\;.\end{equation}

Equations (3.5,6) and (3.8) imply that, for the LC metrics,
\begin{equation} p,\Pi_p,\Pi_{\tilde{A}}\in I\!\!\!\,R\;\;\;\;\;{\rm and}
\;\;\,\;\,\tilde{A}\in I\!\!\!\,R^+\;,\end{equation}
whereas, in the Kasner case,
\begin{equation} p,\Pi_p,\Pi_{\tilde{A}}\in I\!\!\!\,R\;\;\;\;\;{\rm and}
\;\;\,\;\,\tilde{A}\in I\!\!\!\,R^-\;.\end{equation}
Thus, the reduced phase spaces of the two analysed families of solutions
differ
only in the range of values allowed for the canonical variable $\tilde{A}$.
This opens the possibility of simultaneously considering these two kinds of
spacetimes by permitting that $\tilde{A}$ runs over the whole real axis.
In doing this, one is forced to include as acceptable the value
$\tilde{A}=0$, which, from expressions (3.4) and (3.8), can be
reached only in the limit of degenerate metrics. Note, nevertheless, that
from the point of view of the symplectic structure of the reduced phase
space, the inclusion of the surface $\tilde{A}=0$ does not bring on any
singularity.

Admitting then that $\tilde{A}\in I\!\!\!\,R$, the symplectic structure
determined by (3.7) can be identified with that of the cotangent bundle
over $I\!\!\!\,R\times I\!\!\!\,R$. In order to quantize the
system, we can choose the canonical variables $\tilde{A}$,
$\Pi_{\tilde{A}}$, $p$ and $\Pi_p$ as a complete set of real observables,
and represent them as the following operators acting on the vector space of
complex functions $\Psi(\tilde{A},p)$ over $I\!\!\!\,R\times I\!\!\!\,R$:
\begin{equation} \;\hat{\!\!\,\tilde{A}}\Psi(\tilde{A},p)=\;\tilde{A}
\Psi(\tilde{A},p)\;,\;\;\;\;\; \hat{\Pi}_{\tilde{A}}\Psi(\tilde{A},p)
=-i\partial_{\tilde{A}}\Psi(\tilde{A},p)\;,\end{equation}
\begin{equation} \hat{p}\Psi(\tilde{A},p)=\;p
\Psi(\tilde{A},p)\;,\;\;\;\;\; \hat{\Pi}_{p}\Psi(\tilde{A},p)
=-i\partial_{p}\Psi(\tilde{A},p)\;;\end{equation}
whose only non-vanishing commutators are
\begin{equation} \frac{1}{i}[\;\hat{\!\!\tilde{A}},
\hat{\Pi}_{\tilde{A}}]=1\;,
\;\;\;\;\;\;\;\;\frac{1}{i}[\hat{p},\hat{\Pi}_p]=1\;.\end{equation}
The elementary operators (3.11,12) must be self-adjoint, because they
represent real classical observables. These reality conditions select
the physical inner product
\begin{equation} <\Phi,\Psi>=\int_{I\!\!\!\,R\times I\!\!\!\,R}
d\tilde{A}\;dp\;\Phi^{\ast}(\tilde{A},p)\;\Psi(\tilde{A},p)\;.\end{equation}
The Hilbert space of physical states is thus
$L^2(I\!\!\!\,R\times I\!\!\!\,R,d\tilde{A}dp)$.

It is worth remarking that, according to our discussion above [see Eq.
(3.6)], we can interpret the sign of the expectation value of $\tilde{A}$
in any quantum state $\Psi$,
\begin{equation} \eta={\rm sign}\left(\frac{<\Psi,\;\hat{\!\!\,\tilde{A}}
\Psi>}{<\Psi,\Psi>}\right)\;,\end{equation}
as the prediction of a four-dimensional geometry being of either the LC
($\eta=1$) or the Kasner type ($\eta=-1$). Let us suppose then that, eg,
$\eta=-1$ for a particular state $\Psi_{_{0}}$, and that the projection
of $\Psi_{_0}$ on the sector $\tilde{A}>0$,
\begin{equation} \Psi^+_{_0}(\tilde{A},p)=\int_{I\!\!\!\,R^+}dA
\;\delta(\tilde{A}-A)\;\Psi_{_0}(A,p)\;,\end{equation}
is normalizable with respect to the inner product (3.14).
At least formally, one can understand projection (3.16) as the collapse of
the wavefunction $\Psi_{_0}$ produced by a measurement in which one just
detected whether the geometry is static and cylindrically symmetric (ie, of
the LC type). After such an irreversible collapse, it is obvious that
the expectation value of $\tilde{A}$ becomes positive,
so that, from Eq. (3.15), we will have that $\eta=1$ in
the analysed quantum state. In this sense, the collapse of the wavefunction
$\Psi_{_0}$ to $\Psi^+_{_0}$, implied by the physical measurement
explained above, can be interpreted in our quantum theory as leading
to a change in the spatial topology, for the LC and the Kasner
surfaces of constant time possess, respectively, different topologies.

\section{Quantum Gravitational Field of a Cosmic String}
\setcounter{equation}{0}

We have already commented that the LC metrics (1.1) with $p=\frac{1}{2}$
and $C>AB$ represent the exterior gravitational field of a straight cosmic
string. With the aim at constructing a quantum framework for the
analysis of the physical consequences that cosmic strings may imply,
we now want to discuss in detail the quantization of this subfamily
of static and cylindrical solutions.

The collection of metrics that we are going to study can be written in the
form (1.2). We will first show that these metrics can be obtained from the
LC solutions by imposing the following constraints:
\begin{equation} \chi_1\equiv \;p-\frac{1}{2}\;=0\;,\end{equation}
\begin{equation} \chi_2\equiv\;P_p-A(\ln{A}-\ln{D_0}-2p)\;=0\;,
\end{equation}
where $D_0$ is a fixed positive constant, $D_0>0$.

Recalling that the symplectic structure of the reduced phase space of the
LC solutions is given by Eq. (2.9), it is easy to check that the above
equations provide a pair of second-class constraints, with
$\{\chi_1,\chi_2\}=1$. The imposition of (4.1,2) leads to a reduced
model whose only degrees of freedom are the variables $A$ and $\tilde{P}_A$,
the latter defined in Eq. (2.7) [with $|\tilde{C}|$ equal to the parameter
$C$ of Eqs. (1.1,2)]. Making use of Eq. (2.9), it is also
straightforward to see that $A$ commutes with $\chi_1$ and $\chi_2$.
Therefore, after reduction of the system, $A$ and $\tilde{P}_A$ continue to
be canonically conjugate to each other, because their Poisson and Dirac
brackets coincide. Thus, the symplectic form on the physical phase space of
the considered reduced model is
\begin{equation} \Gamma_R=\;dA\wedge d\tilde{P}_A\;.\end{equation}

On the other hand, employing the definiton of $P_p$ in Eq. (2.3) and that
$A>0$, one can prove that constraint (4.2) is equivalent to the condition
\begin{equation} B=\frac{A}{D_0}\;.\end{equation}
Then, provided that constraints (4.1,2) hold, the four-dimensional
metrics (1.1) become
\begin{equation} ds^2=C(-dt^2+d\rho^2)+\frac{A^2}{D_0}\;\rho^2\;
d\phi^2+D_0\;dz^2\;,\end{equation}
with $C,A\in I\!\!\!\,R^+$ and $D_0>0$ fixed. Besides, from Eqs. (2.7) and
(4.4), the variable $\tilde{P}_A$ takes now the expression
\begin{equation} \tilde{P}_A=\frac{1}{2}\;(\ln{A}-\ln{D_0}-\ln{C})\;.
\end{equation}

Finally, to arrive at metrics (1.2), it suffices to perform the following
rescaling in Eq. (4.5)
\begin{equation} \rho^{\prime}=\sqrt{C}\;\rho\;,\;\;\;\;\;\;t^{\prime}
=\sqrt{C}\;t\;,\;\;\;\;\;\;z^{\prime}=\sqrt{D_0}\;z\;.\end{equation}
This rescaling leaves invariant the domains of definition of the radial,
time and axial coordinates, and transforms Eq. (4.5) into
\begin{equation} ds^2=-dt^2+d\rho^2+E^2\;\rho^2d\phi^2+
dz^2\;,\end{equation}
where we have suppressed the primes in the new coordinates and
\begin{equation} E=\;\frac{A}{\sqrt{CD_0}}\;.\end{equation}

If $E\leq 1$, metric (4.8) describes the exterior conical geometry of
a straight cosmic string, with associated deficit angle
equal to [see Eq. (1.3)]
\begin{equation} \Delta=2\pi (1-E)\;.\end{equation}
Thus, for classical solutions which represent the
gravitational field of a cosmic string,
the parameters $A$ and $C$ must satisfy
\begin{equation} A,C\in I\!\!\!\,R^+\;\;\;\;\;\;{\rm and}\;\;\;\;\;\;
\sqrt{CD_0}\geq A\;,\end{equation}
the last inequality guaranteeing that $E\leq 1$. Introducing then the
new variables
\begin{equation} Q=\;A\;\ln{E}\;,\;\;\;\;\;\;\;\;\;\;\;P_Q=\ln{(-\ln{E})}
\;,\end{equation}
where, from Eq. (4.9),
\begin{equation} \ln{E}=\ln{A}-\frac{1}{2}\ln{D_0}-\frac{1}{2}
\ln{C}\;,\end{equation}
a simple computation proves that restrictions (4.11) can be rewritten as
\begin{equation} Q\in I\!\!\!\,R^-\;,\;\;\;\;\;\;\;\;\;\;P_Q\in
I\!\!\!\,R\,.\end{equation}

On the other hand, using Eq. (4.6), the symplectic form (4.3) can be
expressed as
\begin{equation} \Gamma_R=\;dQ\wedge dP_Q\;.\end{equation}
The symplectic structure of the reduced model under consideration can
therefore be identified with that of the cotangent bundle over
$I\!\!\!\,R^-$. Hence, the variables $Q$ and
\begin{equation} L_Q=\;Q\;P_Q\;\end{equation}
provide a complete set of elementary observables. We can represent
these observables as the following operators acting on the vector space
of complex functions over $I\!\!\!\,R^-$:
\begin{equation} \hat{Q}\Psi(Q)=Q\Psi(Q)\;,\;\;\;\;\;\hat{L}_Q
\Psi(Q)=-i\left(Q\partial_Q+\frac{1}{2}\right)\Psi(Q)\;.\end{equation}
The physical inner product is determined by demanding the hermiticity of the
operators $\hat{Q}$ and $\hat{L}_Q$, which represent real classical
observables. This fixes the inner product to be
\begin{equation} <\Phi,\Psi>=\int_{I\!\!\!\,R^-}dQ\;\Phi^{\ast}(Q)\;
\Psi(Q)\;,\end{equation}
so that the Hilbert space of physical states is $L^2(I\!\!\!\,R^-,dQ)$.

To close this section, we will show how one can get predictions for the
observable $E$ from the quantum theory constructed. Notice that, from the
point of view of the classical solutions (4.8), $E$ is the only physically
relevant observable of the conical geometries studied.

{}From the second of relations (4.12), it follows that we can represent the
observable $\ln{(-\ln{E})}$ by the quantum operator $\hat{P}_Q$. Since
the action of $\hat{L}_Q=\widehat{QP_Q}$ has been chosen in Eq. (4.17)
as the symmetric product of $\hat{Q}$ and $-i\partial_Q$, it seems natural
to define the operator $\hat{P}_Q$ by
\begin{equation} \hat{P}_Q\Psi(Q)=-i\partial_Q\Psi(Q)\;.\end{equation}
We conclude in this way that, given any quantum state $\Psi$,
the expectation value
\begin{equation} <\Psi,\hat{P}_Q\Psi>=-i\int_{I\!\!\!\,R^-}dQ\;
\Psi^{\ast}(Q)\;\partial_Q\Psi(Q)\;,\end{equation}
is that corresponding to the observable
\begin{equation} \ln{(-\ln{E})}=\ln{\left(\ln{\left(\frac{2\pi}{2\pi-\Delta}
\right)}\right)}\;,\end{equation}
where we have used Eq. (4.10) to write $E$ in terms of the deficit angle of
the conical geometry.

\section{Conclusions}
\setcounter{equation}{0}

We have succeeded in quantizing the LC family of static and cylindrical
solutions by applying the extended canonical programme proposed by
Ashtekar. In order to achieve this goal, we have first proved that the
reduced phase spaces which cor\-respond, respectively, to the LC
metrics and to the Kasner model have the\linebreak same symplectic structure.
The distinction between these two types of spacetimes amounts to
a different domain of definition for one of the phase space variables
(namely, $P_A$).
Taking this into account, the quantization of the LC solutions
has been easily carried out by following a procedure parallel to that
employed for the Kasner metrics in Ref. 20.

We have also constructed a quantum model which describes the LC and the
Kasner solutions as a whole. This has been possible
because the physical phase spaces of these two types of spacetimes
possess the same symplectic structure and can be joined together
to form a single phase space. The spatial topology of the model
quantized is not uniquely fixed, for the topology of the surfaces
of constant time is different in the LC and in the Kasner spacetimes,
respectively. We have shown that, owing to this fact, some physical
measurements can lead to a change in the spatial topology
predicted by the quantum theory.

Finally, we have quantized the subfamily of LC metrics which provide the
exterior conical geometry of an infinite, straight cosmic string. These
metrics can be obtained from the LC solutions by imposing a pair of
second-class constraints and demanding that the deficit angle of the
four-dimensional geometry be positive. The last requirement guarantees
that the global geometry is conical. In this way, one reaches a reduced
model whose physical phase space can be identified with the cotangent
bundle over $I\!\!\!\,R^-$, and which can be quantized following standard
methods. The corresponding quantum theory allows us to get predictions
about the value of the deficit angle caused by the existence of a
straight cosmic string.
\vspace*{.9cm}

{\bf Acknowledgments}

The author is grateful to P. F. Gonz\'alez D\'{\i}az for helpful
conversations. This work was supported by funds provided by DGICYT
and the Spanish Ministry of Education and Science under Contract
Adjunct to the Project No. PB91-0052.

\end{document}